\documentclass[onecolumn,showpacs,preprintnumbers,amsmath,amssymb]{revtex4}
\usepackage{amsmath,amssymb}
\usepackage{natbib}
\usepackage{url}
\usepackage{graphicx}
\usepackage{psfrag}
\usepackage{color}
\usepackage{epsfig}
\usepackage{capt-of}

\begin{document}

\title{Signatures of the sources in the gravitational waves of a perturbed Schwarzschild black hole}
\author{Juan Carlos Degollado$^{1}$, Dar\'{\i}o N\'u\~nez$^{1}$ and Carlos Palenzuela$^{2}$}

\date{\today}
\label{firstpage}
\affiliation{$^{1}$Instituto de Ciencias Nucleares, Universidad
Nacional Aut\'onoma de M\'exico, Apdo. 70-543, CU, 04510 M\'exico,
D.F., M\'exico.\\
$^{2}$Max-Planck-Institut f\"ur Gravitationsphysik, Albert Einstein Institut,
14476 Golm, Germany} 
\email{jcdegolado@nucleares.unam.mx, nunez@nucleares.unam.mx, carpa@aei.mpg.de}

\begin{abstract}
The explicit form of perturbation equation for the $\Psi_4$ Weyl scalar, containing
the matter source terms, is derived for general type D spacetimes. It is described in detail
the particular case of the Schwarzschild spacetime using in-going penetrating coordinates.
As a practical application, we focused on the emission of gravitational waves when a black hole
is perturbed by a surrounding dust-like fluid matter. The symmetries of the spacetime and
the simplicity of the matter source allow, by means of a spherical harmonic decomposition,
to study the problem by means of a one dimensional numerical code.
\end{abstract}
\maketitle

\section{Introduction}

One of the problems which has received more attention in Numerical Relativity from early
nineties, is the one related to the sources of gravitational waves, which in turn could be observed by the
gravitational wave detectors. The strongest candidates for the current detectors are
the strongly gravitating binary systems, involving highly non-linear evolutions of
either black holes or neutron stars. There are other cases where the geometry of the spacetime
is determined mainly by a big massive compact body which is perturbed by a smaller source.
Typical examples are the Extreme Mass Ratio Inspirals (EMRIs), with small compact objects (ie,
black holes or stars) falling into a supermassive black hole, or the accretion of matter
unto a black hole, where the black hole is surrounded by a much less massive cloud of matter.
The changes in the spacetime by these less massive objects (ie, either smaller black holes
or the accreted disk) can be neglected and the reaction of the central object to the motion
of the matter are well described as perturbations in that background. The perturbative methods
for determining the generation of gravitational waves are expected to correctly describe all
the stages of the process. Such methods, together with a detail analysis of the sources of
the perturbations, are then a promising tool for obtaining the features of the gravitational wave,
with a much less need of computational resources than those needed in the full numerical evolutions.

Our final goal is to accurately describe the gravitational waves generated by the motion
of either small compact bodies (EMRI) or disks of matter in the background of a rotating black
hole using the perturbation theory. Specifically, we work with the curvature perturbations within
the null tetrad formulations developed by Newman and Penrose \cite{NP62}. Basically, some
scalar quantities are constructed by means of projections of the Weyl tensor onto a basis tetrad formed
out from null vectors. Projecting the Bianchi identities on the same null tetrad, we derive the field
equations for those scalar quantities, and then study the perturbed form of these equations in order
to obtain an evolution equation for the perturbed scalar quantity. Usually this scalar quantity
is the perturbed $\Psi_4$ scalar which, as it can be inferred from the ``peeling off theorem''
\cite{Bondi62,Sachs62}, is the one which describes the outgoing gravitational wave.

Since the first derivation of the perturbation equation, \cite{Teu73}, there have been 
several specific works in the same line. Indeed in \cite{Font98}, the authors studied
some of the specific features of the gravitational radiation produced by a 
boosted dust shell surrounding a fixed non-rotating black hole. For the cases considered they needed
to use a 2d code and the spacetime was described with the Boyer-Lindquist coordinates, which introduced several numerical difficulties due to the singular slicing. Other works, as for instance \cite{Kri97, Cam01},
studied the perturbation equation in the case of a rotating black hole, using the so called
penetrating coordinates (also known as Kerr-Schild or Eddington-Finkelstein), avoiding the divergence
problems of the Boyer-Lindquist description, but the authors worked only in the sourceless case.
The use of these penetrating coordinates within the perturbation equation including sources
was started in \cite{MoNun01}, where the authors explicitly derived the equations for several choices of the null tetrad.

In the present work, using the ideas mentioned above, we continue with the description of the gravitational waves including a treatment of the sources which generate them. We rederived the perturbation equations for the gravitational wave, finding a dependence in that equation on the choice of signature. We concentrate on the non rotating black hole case, with dust infalling unto the black hole. Decomposing the dust density in spherical harmonics, we are able to deal with non spherical distributions within a spherical description. We perform the numerical evolution for different initial distributions of the dust and obtain the corresponding gravitational wave, remarking how these waves carry the information of the black hole parameters, in this case the mass, as well as information about the distribution of the source which generated them.

The work is organized as follows: in section \ref{sec:pert_eq}, we present a derivation of the perturbation equation for the perturbed $\Psi_4$ Weyl scalar, including the source terms, for a general type D vacuum background spacetime (the black hole spacetimes belong to this type). Then, in  \ref{sec:Sph_sym}, we use this equation for the static spherical symmetric spacetimes, present the Schwarzschild black hole described in Boyer-Lindquist coordinates as well as in penetrating ones, and explicitly derived the evolution equation for $\Phi=r\,{\Psi_4}$ including sources. In section \ref{sec:sources} we consider that the source term is described by radially infalling dust, make an harmonic decomposition of density, and derive the corresponding equations. We also decompose $\Phi$ in terms of spin weighted spherical harmonics
obtaining a clear correspondence between the modes of matter and the gravitational ones. The perturbation equation can be separated in radial-temporal and angular parts, resulting in a radial-temporal equation for each mode, for which we construct a first order system of equations. In section \ref{sec:evolution} we describe the numerical code used to evolve this system of equations. we first present a well known case to show that the code is working properly, we also analyze the infall of one gaussian pulse and the gravitational response due to the variation in the width of such pulse and we also present the case of three infall pulses of matter and the case of two consecutive pulses varying the initial separation between them. Finally in section  \ref{sec:dis} we present a discussion of the results obtained.

\section{Perturbation equation} \label{sec:pert_eq}

The spinor formulation was introduced by Newman and Penrose \cite{NP62} and used by Teukolsky in \cite{Teu73} to derive a master equation for different fields in a Kerr background. One of the central ideas of such formulation consists in choosing a tetrad of null vectors ${Z_a}^\mu$. These null vectors allow to define directional operators (ie, as the projections of the partial derivative along each null vector) and the covariant derivatives of such null vectors projected on themselves, which plays the role of the Christoffel symbols within this formulation. For a complete review on the subject and its properties, the reader is referred to \cite{Chandra83, Frolov77}. 

Usually the tetrad is defined with the following choice for the the four null vectors: two along the light cone and the other two in the perpendicular plane to the cone. These last two null vectors are usually defined in terms of a complex one and its complex conjugate. The two null vectors along the light cone are taken as real quantities and defined such that they are oriented towards the future and, in the asymptotic region of an hypersurface of constant time, one points outward and the other inward the hypersurface. As it can be seen, there is a large room for several definitions, which might raise some confusion. We make the following choice: we label the two real vectors along the light cone as ${Z_0}^\mu=l^\mu, {Z_1}^\mu=k^\mu$ and choose them both future directed and with $l^\mu$ pointing outward, and $k^\mu$ pointing inward in the asymptotic region of an hypersurface of constant time. This choice  of orientation is consistent with the results obtained for the behavior of the Weyl scalar in the asymptotic regions mentioned in the introduction. The other two null vectors, in the perpendicular plane, will be denoted by a complex vector $m^\mu$ and its conjugate ${m^*}^\mu$, so that ${Z_2}^\mu=m^\mu$ and ${Z_3}^\mu={m^*}^\mu$.

After these choices, there is still the issue of the normalization of the products $l^\mu\,k_\mu$, and $m^\mu\,{m^*}_\mu$. The rest of the products are zero by construction. All the properties mentioned above can be expressed in the following equations:
\begin{equation}
{Z_a}^\mu\,{Z_b}_\mu=\eta_{ab}, \hspace{1cm} g_{\mu\nu}=\eta^{ab}\,Z_{a\,(\mu}\,Z_{b\,\nu)},\label{eqs:tetrad}
\end{equation}
with $\eta_{ab}$ a matrix with the form:
\begin{equation}
\eta_{ab}=\left(\begin{matrix}&0&\, &\eta&\,&0&\,&0&\\&\eta&\,&0&\,&0&\,&0&\\&0&\,&0&\,&0&\,&-\eta&\\&0&\,&0&\,&-\eta&\,&0&\end{matrix}\right),
\end{equation}
with $\eta$ a constant related to the normalization. It is usual to set this constant to one, as it was done in \cite{Chandra83, Frolov77}, and taken from there in \cite{Teu73} to derive the perturbation equation that we mentioned. However, this choice depends on the signature of the spacetime. Indeed, if the spacetime has signature $(+,-,-,-)$, as was the case in the works just mentioned, then this choice is consisten with Eqs.~(\ref{eqs:tetrad}). However, if the spacetime is described using the signature $(-,+,+,+)$, as it is common nowadays, we have to choose $\eta=-1$, in order to remain consistent with Eqs.~(\ref{eqs:tetrad}). Some discussion on the change of the tetrad due to the signature can be found in \cite{Ernst78}.

We are using the definitions for the directional operators, spinor coefficients,
and projections of the Weyl tensor as given in \cite{Chandra83, Frolov77}. In this way, for the directional operators we have:
\begin{equation}
{\bf D}=l^\mu\,\partial_\mu, \hspace{0.5 cm} {\bf \Delta}=k^\mu\,\partial_\mu, \hspace{0.5 cm} {\bf \delta}=m^\mu\,\partial_\mu,
\end{equation}
while the spinor coefficients can be written as:
\begin{eqnarray}
\kappa_s&=&m^\mu\,l_{\mu;\nu}\,l^\nu, \hspace{0.5cm} \tau_s=m^\mu\,l_{\mu;\nu}\,k^\nu, \hspace{0.5cm} \sigma_s=m^\mu\,l_{\mu;\nu}\,m^\nu, \hspace{0.5cm} \rho_s=m^\mu\,l_{\mu;\nu}\,{m^*}^\nu, \nonumber \\
\pi_s&=&k^\mu\,{m^*}_{\mu;\nu}\,l^\nu, \hspace{0.5cm} \nu_s=k^\mu\,{m^*}_{\mu;\nu}\,k^\nu, \hspace{0.5cm} \mu_s=k^\mu\,{m^*}_{\mu;\nu}\,m^\nu, \hspace{0.5cm} \lambda_s=k^\mu\,{m^*}_{\mu;\nu}\,{m^*}^\nu, \nonumber \\
\epsilon_s&=&\frac12\left(k^\mu\,l_{\mu;\nu} + m^\mu\,{m^*}_{\mu;\nu}\right)\,l^\nu, \hspace{0.5cm} \gamma_s=\frac12\left(k^\mu\,l_{\mu;\nu} + m^\mu\,{m^*}_{\mu;\nu}\right)\,k^\nu,\nonumber \\
\beta_s&=&\frac12\left(k^\mu\,l_{\mu;\nu} + m^\mu\,{m^*}_{\mu;\nu}\right)\,m^\nu, \hspace{0.5cm} \alpha_s=\frac12\left(k^\mu\,l_{\mu;\nu} + m^\mu\,{m^*}_{\mu;\nu}\right)\,{m^*}^\nu, \label{eq:coef_spinors}
\end{eqnarray}
where we have added a subindex ${}_s$ to avoid confusion with other symbols which will be defined later in the work. For the projections of the Weyl tensor $C_{\mu\nu\lambda\tau}$, the following five complex quantities, known as Weyl scalars are defined:
\begin{eqnarray}
\Psi_0&=&-C_{\mu\nu\lambda\tau}\,l^\mu\,m^\nu\,l^\lambda\,m^\tau , \nonumber \\
\Psi_1&=&-C_{\mu\nu\lambda\tau}\,l^\mu\,k^\nu\,l^\lambda\,m^\tau , \nonumber \\
\Psi_2&=&-C_{\mu\nu\lambda\tau}\,l^\mu\,m^\nu\,{m^*}^\lambda\,k^\tau , \nonumber \\
\Psi_3&=&-C_{\mu\nu\lambda\tau}\,l^\mu\,k^\nu\,{m^*}^\lambda\,k^\tau , \nonumber \\
\Psi_4&=&-C_{\mu\nu\lambda\tau}\,k^\mu\,{m^*}^\nu\,k^\lambda\,{m^*}^\tau. \label{eq:Psis}
\end{eqnarray}

In order to obtain the perturbed equations, one starts from the Bianchi identities and the definition of the Riemann tensor, namely
\begin{equation}
R_{\mu\nu\lambda\tau\,;\sigma} + R_{\mu\nu\sigma\lambda\,;\tau} + R_{\mu\nu\tau\sigma\,;\lambda} =0~,
\hspace{1cm} R_{\sigma\mu\nu\lambda}\,{Z_a}^\sigma=Z_{a\,\mu;\nu\lambda}-Z_{a\,\mu;\lambda\nu},\label{eqs:ByR}
~~.
\end{equation}
These equations can be projected on the null tetrad, leaving unspecified the normalization constant $\eta$. With the above definition (\ref{eq:coef_spinors},\ref{eq:Psis}), the equations (\ref{eqs:ByR}) translate in equations for the Weyl scalars and spinor coefficients \cite{Chandra83}. The perturbed
expression of these equations is computed for the case of vacuum type D spacetimes to obtain a master
equation for the perturbed Weyl scalar $\Psi_4$ including the source terms \cite{Teu73}:
\begin{eqnarray}
&&[\left({\bf \Delta} + \eta\, \left(4\,\mu_s + {\mu_s}^* + 3\,\gamma_s -{\gamma_s}^* \right)\right)\,\left({\bf D} - \eta\,\left(\rho_s -4\,\epsilon_s \right) \right) - \left({\bf \delta^*} + \eta\, \left(3\,\alpha_s + {\beta_s}^* + 4\,\pi_s - {\tau_s}^* \right)\right)\,\left( {\bf \delta} + \eta\,\left(4\,\beta_s-\tau_s \right) \right) 
\nonumber \\
&& - 3\, \eta \, \Psi_2 ] \, {\Psi_4}^{(1)} = \eta \,\frac{K} 2\,T_4 ~~. \label{eq:PertPsi4f}
\end{eqnarray}
with $K=8\,\pi$ in geometrized units. The source terms, denoted by $T_4$, are given as follows:
\begin{equation}
T_4 ={\cal{{\hat T}}}^{k\,k}\,{T}_{k\,k} + {\cal{{\hat T}}}^{k\,m^*}\,{T}_{k\,m^*} 
    + {\cal{{\hat T}}}^{m^*\,m^*}\,{T}_{m^*\,m^*}, \label{eq:T4g}
\end{equation}
where we have introduced the following operators ${\cal{{\hat T}}}^{ab}$ acting on the projections of the
perturbed source term ${T}_{\mu\,\nu}$ on the null tetrad,
\begin{eqnarray}
{\cal{{\hat T}}}^{k\,k}&=&-\left({\bf \delta^*} + \eta\, \left(3\,\alpha_s + {\beta_s}^* + 4\,\pi_s - {\tau_s}^* \right)\right)\,\left({\bf \delta^*} +\eta\,\left(2\,\alpha_s + 2\,{\beta_s}^* - {\tau_s}^* \right)\right), \nonumber \\
{\cal{{\hat T}}}^{k\,m^*}&=&\left({\bf \Delta} + \eta\, \left(4\,\mu_s + {\mu_s}^* + 3\,\gamma_s -{\gamma_s}^* \right)\right)\, \left({\bf \delta^*}+2\,\eta\,\left(\alpha_s -{\tau_s}^* \right)\right) + \nonumber \\
&& \left({\bf \delta^*} + \eta\, \left(3\,\alpha_s + {\beta_s}^* + 4\,\pi_s - {\tau_s}^* \right)\right)\,\left({\bf \Delta} +  2\eta\,\left({\mu_s}^* + \gamma_s \right)\right), \nonumber \\
{\cal{{\hat T}}}^{m^*\,m^*}&=&-\left({\bf \Delta} + \eta\, \left(4\,\mu_s + {\mu_s}^* + 3\,\gamma_s -{\gamma_s}^* \right)\right)\,\left({\bf \Delta} +\eta\,\left({\mu_s}^* + 2\,\gamma_s - 2\,{\gamma_s}^*\right)\right). \label{ops:mat_gen}
\end{eqnarray}
By choosing $\eta=1$, we recover the perturbation equation for $\Psi_4$ derived by Teukolsky \cite{Teu73} by using the signature $(+,-,-,-)$. However, for the other choice of signature, we must take $\eta=-1$ in order to have a consistent description of the gravitational perturbation. We remark that the changes in the perturbation equation due to the choice of signature are very noticeable and important at this level of the perturbation equation. However, for a specific case (ie, for a given spacetime with a given selection of the signature), the final perturbation equation in a explicit coordinate system is invariant with respect to the choice of signature. It is only at the level of the perturbation equation within the null tetrad formulation, that care must be taken with respect of the election of signature made.

\section{Spherical symmetric spacetimes} \label{sec:Sph_sym}

In this section we study the perturbation equation (\ref{eq:PertPsi4f}), with the sources
given by eqs.~(\ref{eq:T4g},\ref{ops:mat_gen}), for a static spherically symmetric space time.
In this case, the most general line element can be written as:
\begin{equation}
ds^2=-\left(\alpha^2 - \gamma^2\,\beta^2\right)\,dt^2 + 2\,\gamma^2\,\beta\,dt\,dr\, + \gamma^2\,dr^2 + r^2\,d\Omega^2, \label{eq:lel_esfgen}
\end{equation}
where $d\Omega^2=d\theta^2 + \sin^2\theta\,d\varphi^2$ is the solid angle element, and the lapse, $\alpha$,
the radial component of the shift vector, $\beta$, and the $rr$-component of the metric, $\gamma^2$, are
functions of the $r$ coordinate only. Notice that we have already chosen our signature, so that in the perturbation
equation we have to use $\eta=-1$.

The next step is to construct the null tetrad for this spacetime by using eq.~(\ref{eqs:tetrad}) with
$\eta=-1$. With respect to the two real null vectors, we choose them to describe the temporal-radial part of the spacetime, so that they only have $t$ and $r$ components. We define both of them to point to the future in the asymptotic flat region, and in a timelike hypersurface $l^\mu$ will point outward while $k^\mu$ will point inwards. Under these conditions, there is left only one
free function, which we denote by $k_0=k_0(r)$. The remaining two null vectors have now to describe the angular part. After the normalization conditions are imposed on them, there is left another free function which describes a rotation in the angular plane. We have choosen this function to be equal to one, since other choices have no simplifying consequences on the final perturbation equation.
In this way, our general null tetrad for the line element (\ref{eq:lel_esfgen}) reads as:
\begin{equation}
l^\mu=\frac1{2\,\alpha^2\,k_0}\left(1,\frac{\alpha}{\gamma}-\beta,0,0\right)~,
\hspace{0.5cm} k^\mu=k_0\,\left(1,-\frac{\alpha}{\gamma}-\beta,0,0\right)~,
\hspace{0.5cm} m^\mu=\frac1{\sqrt{2}\,r}\left(0,0,1,i\,\csc\theta\right).
\label{tet_esfgen}
\end{equation}
Within this tetrad, the only non-zero spinor coefficients and Weyl scalars, given by eq. ~(\ref{eq:coef_spinors}), are
\begin{eqnarray}
\rho_s=\frac{\frac{\alpha}{\gamma}-\beta}{2\,r\,\alpha^2\,k_0}~, \hspace{0.5cm} 
\mu_s= \frac{k_0\, \left(\frac{\alpha}{\gamma} + \beta \right)}{r}~, \hspace{0.5cm} 
\epsilon_s=-\frac{ \partial_r \left( r\, \alpha \, \gamma \, \rho_s \right)}{2 \, \alpha \,\gamma}~,
\nonumber \\
\gamma_s=-\frac{ \partial_r \left(r\,\alpha\,\gamma\,\mu_s\right)}{2\,\alpha\,\gamma}~,
\hspace{0.5cm} 
\beta_s=\frac{\cot\theta}{2\,\sqrt{2}\,r}~, \hspace{0.5cm} 
\alpha_s=-\beta_s~, \hspace{0.5cm} 
\Psi_2=\frac{\partial_r \left(r^2\,\alpha\,\gamma\,\rho_s\,\mu_s\right)}{r\,\alpha\,\gamma}.
\label{coef:spin_esfgen}
\end{eqnarray}
As mentioned in the introduction, within the spherical symmetric spacetimes, the perturbation equation ~(\ref{eq:PertPsi4f}) has been studied to describe the gravitational waves of a non rotating black hole.
Among these studies, there are those which have used the Boyer Lindquist coordinate system including sources,
and those using penetrating coordinates in the sourceless case. Let us give a brief review of these works.

In \cite{Font98}, the authors made pioneering numerical studies of the matter perturbing a Schwarzschild black hole 
described in the Boyer-Lindquist coordinates, which are obtained from the general line element (\ref{eq:lel_esfgen}), with
\begin{equation}
\alpha^2=1-2\,\frac{M}r~, \hspace{0.5cm} \gamma^2=\frac1{\alpha^2}~, \hspace{0.5cm} \beta=0 ~~.
\label{coef:met_BL}
\end{equation}
In that work they used the Kinnersley tetrad, which can be obtained from our general tetrad form (\ref{tet_esfgen}) with the choice of metric coefficients (\ref{coef:met_BL}) and $k_0=\frac12$. A direct substitution of these metric coefficients and tetrad vectors on Eqs.~(\ref{coef:spin_esfgen}, \ref{eq:PertPsi4f}), allows us to obtain the spinor coefficients and then the following  perturbation equation:
\begin{eqnarray}
&&\left[-\frac{r^3}{r-2\,M}\frac{\partial^2}{\partial t^2} +  r\,\left(r-2\,M\right)\frac{\partial^2}{\partial r^2} +  4\,\frac{r\,\left(r-3\,M\right)}{r-2\,M}\,\frac{\partial}{\partial t} + 2\,\left(3\,r-7\,M\right)\,\frac{\partial}{\partial r} +
4\,\left(1-4\,\frac{M}{r}\right) + \right.\nonumber\\ 
&& \left.\frac{\partial^2}{\partial \theta^2} + \frac1{\sin^2\theta}\,\frac{\partial^2}{\partial \varphi^2} + \cot\theta\,\frac{\partial}{\partial \theta} - 4i\frac{\cos\theta}{\sin^2\theta}\,\frac{\partial}{\partial \varphi}  - 2\frac{\cos^2\theta + 1}{\sin^2\theta} \right]\,{\Psi_4}^{(1)}=2\, K \,r^2\,T_4. \label{eq:PertPsi4_BL}
\end{eqnarray}
This equation is equivalent in the non-rotating case to the one used in \cite{Font98}, as well as the one obtained by Teukolsky \cite{Teu73}, for the non-rotating case. We do not follow such derivations because to our purpose this equation is enough. Indeed, we confirm that there is a divergence problem at the horizon (ie, $r=2\,M$), as expected from the Boyer-Lindquist coordinates. This coordinate singularity forces the introduction of the tortoise coordinate in order to overcome this divergence. Also, there appears a remarkable problem outside the horizon at $r=3\,M$; there is a change of sign in the coefficient of the first temporal derivative. The authors in \cite{Font98} needed to include dissipation techniques to prevent the appearance of numerical instabilities at this point. The coefficients of the source terms also present these kind of coordinates problems and was needed the ingenuity of the authors to perform the numerical evolutions and to be able to study the gravitational answer of the black hole to the falling of several cases of dust shells studied in that work. We consider interesting to remark that the signature chosen in these two works \cite{Font98,Teu73} was $(+,-,-,-)$ . As mentioned above, when the derivation is done consistently, the final perturbation equation (ie, the eq.~(\ref{eq:PertPsi4_BL}) in this case), is independent of the choice of signature. 

There are also works where the perturbation equation has been numerically solved even for the
case of a rotating black hole, either by using Boyer-Lindquist coordinates \cite{Kri97} and
facing similar problems as those mentioned above, or by using a coordinate system where the metric
coefficients are free of coordinate singularities in the radial sector \cite{Cam01}. In this case, the perturbation equation is also free of coordinates singularities and the inner boundary can be set to be inside the black hole horizon, allowing the use of the excision technique to avoid dealing with the physical singularity. This description results in a system of equations whose numerical implementation is much simpler that when using Boyer-Lindquist coordinates. Both of these works considered the
vacuum spacetimes, studying the evolution of given initial configurations of the perturbed Weyl scalar ${\Psi_4}$.

It seems clear that the best option is to continue these works by considering realistic sources while
using the coordinate description free of singularities . In the rest of the present work we perform such program for the Schwarzschild case, working along the lines initiated by Papadopoulos and Font \cite{Font98}, but in penetrating coordinates. The line element of the Schwarzschild black hole,
written by using the ingoing Kerr-Schild coordinates, has the form:
\begin{equation}
ds^2=-\left(1-\frac{2 M}{r} \right)\,dt^2 + \frac{4 M}{r}\,dt\,dr + \left(1+ \frac{2 M}{r} \right)\,dr^2 + r^2\,d\Omega^2 ~~.
\label{eq:lel_KS}
\end{equation}
where, by comparison with the general line element (\ref{eq:lel_esfgen}), we see that the lapse, the 
radial component of the shift function and the $rr$-component of the metric are given by
\begin{equation}
\gamma^2=1+\frac{2 M}r~, \hspace{0.5cm} 
\beta=\frac{2 M}{r\,\gamma^2}~, \hspace{0.5cm} 
\alpha^2=\frac{1}{\gamma^2}~~,  \label{coef:met_KS}
\end{equation}
so all of them are regular outside the singularity. By inspecting the expression (\ref{tet_esfgen}) for the tetrad in the general spheric case becomes clear that $k_0=1$ gives a particular simple description. This choice differs with the one chosen in \cite{Cam01}, which for the Schwarzschild case reduces to consider $k_0=\frac1{r^2}$. The explicit form of our tetrad in these coordinates is given by:
\begin{equation}
l^\mu=\frac12\,\left(1+\frac{2 M}{r},1-\frac{2 M}{r},0,0\right)~, \hspace{1cm}  k^\mu=\left(1,-1,0,0\right)~, 
\hspace{1cm} m^\mu=\frac1{\sqrt{2}\,r}\left(0,0,1,i\,\csc\theta\right)~,
\end{equation}
as well as the non-zero spinor coefficients and Weyl scalar: 
\begin{eqnarray}
&\mu_s&=\frac{1}{r}~, \hspace{0.5cm} \rho_s=\frac{r-2\,M}{2\,r^2}~, \hspace{0.5cm}  \epsilon_s=-\frac{M}{2\,r^2}~,
\nonumber \\
&\alpha_s&=\frac{\cot\theta}{2\,\sqrt{2}\,r}~, \hspace{0.5cm} \beta_s=-\alpha_s~,
\hspace{0.5cm} \Psi_2=\frac{M}{r^3}~.
\end{eqnarray}
A straightforward substitution of these quantities in the perturbation equation leads to the final
equation for the perturbed scalar of ${\Psi_4}$:
\begin{equation}
\left[\square^{\Psi}_{tr} + \square_{\theta\,\phi} \right]\,{\Psi_4}^{(1)}=2\, K\, r^2\,{T_4}~. \label{eq:pertPsi4KS}
\end{equation}
where
\begin{eqnarray}
\square^{\Psi}_{tr}&=&-\left(r^2+2\,M\,r\right)\,\frac{\partial^2}{\partial t^2} +  \left(r^2-2\,M\,r\right)\,
\frac{\partial^2}{\partial r^2} + 4\,M\,r\,\,\frac{\partial^2}{\partial t \partial r} + 2\,\left(2\,r+3\,M\right)\,\frac{\partial}{\partial t} + 6\,\left(r - M\right)\,\frac{\partial}{\partial r} + 4 \label{op:Pertrt} \\
\square_{\theta\varphi}&=& \frac{\partial^2}{\partial \theta^2} + \frac1{\sin^2\theta}\,\frac{\partial^2}{\partial \varphi^2} + \cot\theta\,\frac{\partial}{\partial \theta} - 4\,i\,\frac{\cos\theta}{\sin^2\theta}\,\frac{\partial}{\partial \varphi} - 2\,\frac{1+\cos^2\theta}{\sin^2\theta}~. \label{op:Pertthph}
\end{eqnarray}
On the other side, the explicit form of the operators defined by the eqs.~(\ref{eq:T4g},\ref{ops:mat_gen})
for the source term $T_4$ take the form:
\begin{eqnarray}
{\cal{{\hat T}}}^{k\,k}&=&-\frac1{2\,r^2}\,{\bar\eth}_{-1}\,{\bar\eth}_0~, \label{op_tkk}\\
{\cal{{\hat T}}}^{k\,m^*}&=&-\frac{\sqrt{2}}{r}\,\left(\frac{\partial}{\partial t}-\frac{\partial}{\partial r} -\frac3r\right){\bar\eth}_{-1}~,\\
{\cal{{\hat T}}}^{m^*\,m^*}&=&-\left(\frac{\partial^2}{\partial t^2}-2\,\frac{\partial^2}{\partial t \partial r} + \frac{\partial^2}{\partial r^2} -\frac6r\left(\frac{\partial}{\partial t}-\frac{\partial}{\partial r}\right) +\frac4{r^2}\right)~, \label{op_tmm}
\end{eqnarray}
where we used the ``eth'' and ``eth-bar'' \cite{Newm66,Gold67}, defined as:
\begin{eqnarray}
\eth_s=-\left(\frac{\partial}{\partial \theta} + i\,\csc\theta\,\frac{\partial}{\partial \varphi} - s\,\cot\theta\right)\equiv \eth_0 + s\,\cot\theta~, \label{op:eth} \\
{\bar\eth}_s=-\left(\frac{\partial}{\partial \theta} - i\,\csc\theta\,\frac{\partial}{\partial \varphi} + s\,\cot\theta\right)\equiv {\bar\eth}_0 - s\,\cot\theta~, \label{op:beth}
\end{eqnarray}
in which the subindex $s$ indicates that they act on a quantity of spin weight $s$. These operators, inspired on the angular operators of quantum mechanics, were designed to act on the spin weighted spherical harmonics ${Y_s}^{l,m}(\theta,\varphi)$. The spherical harmonics live on the sphere, where 
they form a functional basis, and are defined only for $|s|\leq l$.  When the $\eth$ operator acts on the spin weighted spherical harmonic, it raises the spin weight of the harmonic:
\begin{equation}
\eth_s\,{Y_s}^{l,m}=\sqrt{\left(l-s\right)\,\left(l+s+1\right)}\,{Y_{s+1}}^{l,m}~, \label{op:eth_Y}
\end{equation}
and, when the ${\bar \eth}$ operator act on the spin weighted spherical harmonic, it lowers the spin weight of the harmonic:
\begin{equation}
{\bar \eth}_s\,{Y_s}^{l,m}=-\sqrt{\left(l+s\right)\,\left(l-s+1\right)}\,{Y_{s-1}}^{l,m}~, \label{op:beth_Y}
\end{equation}
so the operator $\eth$ is also called spin raising and ${\bar \eth}$ spin lowering. For a larger discussion on these harmonics and on how to use them to extract physical information carried by the gravitational wave see for instance \cite{Miguel3p1,Ruiz08}. 
For us, it is more suitable to work with the function $\Phi=r\,\,{\Psi_4}$, which is expected to have a constant behavior in the regions far from the black hole due to the peeling theorem ( see \cite{Bondi62,Sachs62}), so that the final perturbation equation becomes:
\begin{equation}
\left[{\square}^{\Phi}_{tr} + \square_{\theta\,\phi} \right]\,\Phi=2\,K\,r^3\,{T_4}~.
\label{eq:pertPhi1}
\end{equation}
where there was only a change in the temporal radial operator, now taking the explicit form:
\begin{equation}
{\square}^{\Phi}_{tr}=-\left(r^2+2\,M\,r\right)\,\frac{\partial^2}{\partial t^2} +  \left(r^2-2\,M\,r\right)\,\frac{\partial^2}{\partial r^2} + 4\,M\,r\,\,\frac{\partial^2}{\partial t \partial r} + 2\,\left(2\,r+ M\right)\,\frac{\partial}{\partial t} + 2\,\left(2\,r - M\right)\,\frac{\partial}{\partial r} + 2\frac{M}r~.
\label{op:Pert1rt}
\end{equation}
This equation Eq.~(\ref{eq:pertPhi1}), together with the definitions of the operators 
(\ref{op_tkk}-\ref{op_tmm}), is the equation for the gravitational perturbation of a Schwarzschild black hole
in Kerr-Schild coordinates due to an arbitrary test source $T_{\mu\nu}$. There are several interesting features worth noticing before proceeding further. First, that the equation (\ref{op:Pert1rt}) is regular in all the domain of interest due to the choice of smooth penetrating horizon coordinates. Second, there is a complete splitting between the radial-temporal part and the angular one, with a principal part identical to the corresponding for the wave equation. This implies that the gravitational character of the equation is determined by the lower order terms (ie, first order derivatives and independent ones). The hyperbolicity of the gravitational perturbation equation (\ref{eq:pertPhi1}) is guarantied by the hyperbolicity of the scalar wave, which was shown for instance in ref. \cite{lLmSlKrOoR06}. Finally, it is remarkable the simple form of the operators (\ref{op_tkk}-\ref{op_tmm}), which have a different angular action on the corresponding projections of the perturbed source term. This suggests that,
at least in some cases, there may be a preferred angular decomposition of such components. In the next section we will present a simple example that will show this feature.

\section{Decoupling the angular part for dust matter}\label{sec:sources}

Let us consider the matter source to be described by a dust-like fluid, that is
\begin{equation}
{T}_{\mu\nu}=\rho\,u_\mu\,u_\nu~,
\end{equation}
where $\rho$ is the rest mass density and $u_\mu$ the four velocity of the dust. Furthermore, we will
consider that the fluid is infalling radially in the black hole, so this four
velocity has only temporal and radial components,
\begin{equation}\label{def:four_vel_fluid}
u^\mu=\left(u^0,u^1,0,0\right)~,
\end{equation}
which are functions of $r$ and $t$ only.
The evolution of the fluid is described by the continuity equation for the current vector, $J^\mu=\rho\,u^\mu$, 
and the conservation equation for the stress energy tensor:
\begin{eqnarray}
{J^\mu}_{;\mu}&=&0~, \label{eq:cont}\\
{T^{\nu\mu}}_{;\mu}&=&0~. \label{eq:divtmunu}
\end{eqnarray}
Althought the equations (\ref{eq:divtmunu}) imply in general the Euler equations, in the case of dust
they simply reduce to the geodesic motion $u^\nu\,u_{\mu ; \nu}=0$. These geodesic equations,
by using the symmetries of the spacetime and the normalization on the velocity $u^\mu\,u_\mu=-1$, can be integrated once, so the four velocity can be expressed in terms of the constants of motion and the position of the particle \footnote[1]{Considering the constraint on the velocity as the Lagragian, ${\cal L}=g_{\mu\,\nu}\,u^\mu\,u^\nu=-1$, as long as the time is a cyclic coordinate, we obtain the conserved quantity: $E=-\frac12\,\frac{\partial {\cal L}}{\partial u^0}$}. For the radial motion that we are considering, we obtain the following expressions for the components of the four velocity:
\begin{equation}
u^1=\pm\,\sqrt{E^2 - 1 + 2\frac{M}r}~, \hspace{1cm} u^0=\frac{E\,r + 2\,M\,u^1}{r-2\,M}~, \label{eqs:us}
\end{equation}
which determine the components of the velocity of each particle of the fluid at a given position and time. In these last equations $E$ is a constant of motion along the trajectory, associated with the energy, and we will consider the minus sign for the $u^1$ component, as we will study a shell initially already infalling. Subsequently, the hydrodynamical problem reduces to the continuity equation (\ref{eq:cont}), which provides a evolution equation for the density, namely:
\begin{equation}
\partial_t\,(\sqrt{-g} \rho u^0) + \partial_r (\sqrt{-g} \rho u^1)=0~.
\label{eq:evolrho0} 
\end{equation}
where $\sqrt{-g}$ is the determinant of the four-dimensional metric.

Let us now study more in detail the source terms appearing in the perturbation equation. Within
the simple form of the four-velocity (\ref{def:four_vel_fluid}), only the projections of the stress energy tensor along the light cone in the k-direction will not vanish, that is
\begin{equation}
  {T}_{k\,m^*}={T}_{m^*\,m^*}=0~,~~~~
  {T}_{k\,k}=\left(k^\mu\,u_\mu\right)^2\,\rho=(u^0+u^1)^2\,\rho ~.
\end{equation}
In this simplified case, the source terms for the perturbation equation (\ref{eq:pertPhi1})
are just given by
\begin{equation}
  {T}_{4}=  {\cal{{\hat T}}}^{k\,k} T_{k\,k} 
   = -\frac{(u^0+u^1)^2}{2\,r^2}\,{\bar\eth}_{-1}\,{\bar\eth}_0 \,\rho~.
  \label{op_T4_fin}\\
\end{equation}
This simple expression indicates that the angular part can be decoupled in the source terms by
decomposing the density in terms of the usual spherical harmonics (with zero weight), that is,
\begin{equation}
\rho=\sum\limits_{lm}\,\rho_{l,m}(t,r)\,{Y_0}^{l,m}(\theta,\phi)~. 
\label{dec:rho_lm}
\end{equation}
The action of the bar eth operators (\ref{op:beth}) acting on an harmonic of spin zero 
lowers twice the spin weight to $-2$,
\begin{equation}
\bar{\eth}_{-1}\,\bar{\eth}_{0}\,{Y_0}^{l,m}=-\sqrt{l\,\left(l+1\right)}\,\bar{\eth}_{-1}\,{Y_{-1}}^{l,m}=\sqrt{\left(l-1\right)\,l\,\left(l+1\right)\,\left(l+2\right)}\,{Y_{-2}}^{l,m}~.
\end{equation}
Collecting these results we get that the source term for this dust like case has the form
\begin{equation}
T_4={\cal {\hat T}}^{k\,k}\,{T}_{k\,k}= -\frac{(u^0+u^1)^2}{2\,r^2}\sum\limits_{lm}\,\rho_{l,m}(t,r)\,\sqrt{\left(l-1\right)\,l\,\left(l+1\right)\,\left(l+2\right)}\,{Y_{-2}}^{l,m}~.
\label{eq:T4_a}
\end{equation}
Notice that we are able to describe configurations of dust not necessarily spherically symmetric.
The mass of the cloud will be given by the $(l,m)=(0,0)$ mode, which do not generate any gravitational
perturbation, while the other modes will describe the over and under densities to this spheric base mode. 

The same decoupling of the angular part can be performed to the gravitational perturbation equation, by decomposing the Weyl scalar $\Phi$ as a function of spherical harmonics
with spin weight $-2$ \cite{Chandra83}:
\begin{equation}
\Phi=\sum\limits_{lm}\,R_{l,m}(t,r)\,{Y_{-2}}^{l,m}(\theta,\phi)~,
\label{def:Phi_lm}
\end{equation}
such that the angular operator for the gravitational perturbation equation (\ref{op:Pertthph})
is expressed in terms of the eth operators defined in Eqs.~(\ref{op:eth},\ref{op:beth}) in the following way:
\begin{equation}
\square_{\theta\varphi}={\bar\eth}_{-1}\,\eth_{-2}~.
\end{equation}
Using the properties of the eth operators acting on the harmonics, given by eqs.~(\ref{op:eth_Y}, \ref{op:beth_Y}), it can be seen that the ${Y_{-2}}^{l,m}$ are eigenfunctions of the angular operator:
\begin{equation}
\square_{\theta\varphi}\,{Y_{-2}}^{l,m}={\bar\eth}_{-1}\,\eth_{-2}\,{Y_{-2}}^{l,m}= -\left(l-1\right)\,\left(l+2\right)\,{Y_{-2}}^{l,m}~.
\label{eq:Ym2}
\end{equation}

Collecting these results we get that all the terms in the perturbation equation, including the sources, are multiples of ${Y_{-2}}^{l,m}$, so that we can obtain an equation for each mode $(l,m)$ of the radial-temporal part of the $\Phi$:
\begin{eqnarray}
&&-\left(r^2+2\,M\,r\right)\,\frac{\partial^2 R_{l,m}}{\partial t^2} +  \left(r^2-2\,M\,r\right)\,\frac{\partial^2 R_{l,m}}{\partial r^2} + 4\,M\,r\,\frac{\partial^2 R_{l,m}}{\partial t \partial r} + 2\,\left(2\,r+M\right)\,\frac{\partial R_{l,m}}{\partial t} + 2\,\left(2\,r-M\right)\,\frac{\partial R_{l,m}}{\partial r} + \nonumber\\
&& \left(2\frac{M}r-\left(l-1\right)\,\left(l+2\right)\right)\,R_{l,m} + 4\,\pi\,r^3\,\left(\frac{E - \sqrt{E^2-1+2\frac{M}r}}{r-2\,M}\right)^2\,\sqrt{\left(l-1\right)\,l\,\left(l+1\right)\,\left(l+2\right)}\,\rho_{l,m}=0~. \label{eq:pert_f}
\end{eqnarray}
An analogous expression can be found for the density modes from eq. (\ref{eq:evolrho0}), where
after substituting the decomposition (\ref{dec:rho_lm}) and the explicit expressions of the four-velocity
(\ref{eqs:us}), one obtains
5
\begin{equation}
\partial_t\,\rho_{l,m} + v^r\,\partial_r\,\rho_{l,m}+2\frac{E^2-1+\frac{3\,M}{2\,r}}{r\,\left(E^2-1+\frac{2\,M}{r}\right)}\,v^r\,\rho_{l,m}=0~, \label{eq:evolrho}
\end{equation}
where we have introduced the coordinate velocity, $v^r=\frac{dr}{dt}=\frac{u^1}{u^0}$.
It can be shown that all the terms are regular and well behaved in all the domain outside $r=0$. Any initial arbitrary matter distribution can be expanded in terms of spherical harmonics with spin weight $s=-2$, 
althought neither the monopole nor the dipole modes will generate any gravitational reaction. This decomposition allow us to study, with a 1D numerical code, any radially in-falling dust matter distribution and its gravitational reaction. Notice also that there is a degeneracy with respect
to the $m$ modes, since only the label $l$ appears in the evolution equation (\ref{eq:pert_f}).
This is due to the strong symmetry conditions; the background
is spherically symmetric and we are restricting the fluid movement to be radially infalling. Indeed, the $m$ modes give a description of the functions in the azimuthal angle $\varphi$, and due to the
spherical symmetry of the background and of the motion of the fluid, we see that the action of the source on the black hole is independent of the azimuthal angle of the source.

Thus, in this simple case of radial infall, each mode of matter awakes the corresponding mode in the gravitational radiation. In order to study the dependence of the gravitational wave on the initial distribution of the source which generates it, we will solve numerically the gravitational perturbation equation by recasting it as a system of first order evolution equations. In the next section we describe our numerical procedure to solve this system of equations and the results obtained from these evolutions.

\section{Numerical evolution}\label{sec:evolution}

The Teukolsky equation (\ref{eq:pert_f}) can be reduce to a first order system of evolution equations
by following a standard procedure used for the scalar wave equation \cite{Zeng08}, which guaranties a well posed system of equations. We define the functions $\psi_{l,m}$ and $\Pi_{l,m}$ in terms of the $R_{l,m}$ functions as 
\begin{equation}
\psi_{l,m}=\partial_r\,R_{l,m}~, \hspace{1cm}
\Pi_{l,m}=\frac{\sqrt{\gamma_{rr}}}{\alpha}\,\left(\partial_t\,R_{l,m} -\beta^r\,\psi_{l,m} \right)~, \label{eqs:psi_pi_g}
\end{equation}
so, using the lapse, shift and metric functions given in Eq.~(\ref{coef:met_KS}) for the Schwarzschild spacetime described in penetrating coordinates, we get for the definition of $\Pi_{l,m}$
\begin{equation}
\Pi_{l,m}=\frac{r+2\,M}r\,\partial_t\,R_{l,m} -2\,\frac{M}r\,\psi_{l,m}~, \label{eq:Pi}
\end{equation}
and after a straightforward derivation for the evolution equation for $\Pi_{l,m}$, derived from Eq.(\ref{eq:Pi},\ref{eq:pert_f}), we get the following first order system of equations:
\begin{eqnarray}
\partial_t\,R_{l,m}&=&\frac{1}{r+2\,M}\left(r\,\Pi_{l,m} + 2\,M\,\psi_{l,m} \right)~,
\label{eq:evolR}\\
\partial_t\,\psi_{l,m}&=&\partial_r\left(\frac{1}{r+2\,M}\left(r\,\Pi_{l,m} + 2\,M\,\psi_{l,m} \right)\right)=\frac{1}{r+2\,M}\left(r\,\partial_r\,\Pi_{l,m} + 2\,M\,\partial_r\,\psi_{l,m} \right) + \frac{2\,M}{(r+2\,M)^2}\left(\Pi_{l,m} - \psi_{l,m} \right)~,
\label{eq:evolPsi}\\
\partial_t\,\Pi_{l,m}&=&\frac{1}{r+2\,M}\left(2\,M\,\partial_r\,\Pi_{l,m} + r\,\partial_r\,\psi_{l,m} \right) + \frac{2}{r\,(r+2\,M)^2}\left(\left(2\,r^2+5\,M\,r+4\,M^2\right)\,\Pi_{l,m} + \left(r+4\,M\right)\,\left(2\,r+3\,M\right)\psi_{l,m} \right) \nonumber \\
&&+ \left(2\frac{M}{r^3} - \frac{\left(l-1\right)\,\left(l+2\right)}{r^2}\right)\,R_{l,m} + 4\,\pi\,r\,\left(\frac{E - \sqrt{E^2-1+2\frac{M}r}}{r-2\,M}\right)^2\,
\sqrt{\left(l-1\right)\,l\,\left(l+1\right)\,\left(l+2\right)}\,\rho_{l,m}~,
\label{eq:evolPi}
\end{eqnarray}
This system will be evolved simultaneously with the continuity equation ~(\ref{eq:evolrho}) for the
density of the source by using the Method Of Lines (MoL) \cite{MoL}. Within this approach, the space derivatives are first discretized in order to obtain a semi-discrete system. The semi-discrete system is
evolved in time by using (at least) a third order Runge-Kutta. Due to the simplicity of our equations and the absence of shock, standard fourth order finite difference operators have been used for discretizing
the space derivatives. In addition, a small Kreiss-Oliger dissipation has been added
in order to damp the unphysical high-frequency modes.

Our computational grid has some boundaries at finite positions. In general, data at 
those boundaries has to be provided in order to get a stable evolution. In our case, the
boundaries are located at $r_{in}= M$ and $r_{out}$ varies according to the case analyzed. The interior boundary $r_{in}$
is placed inside the apparent horizon of the black hole. In this case, it is not necessary to impose a boundary condition, since all the characteristic fall into the black hole; all the information travels towards the same direction. The outer boundary $r_{out}$ is placed far away from the black hole,
and we are imposing outgoing boundary conditions for the fluid density $\rho$ and maximally
dissipative BC for the $\Phi$.

We have consider as a initial data a shell of matter which can be written simply as
the superposition of two spherical harmonic modes, namely
\begin{equation}\label{general_rho}
   \rho = \rho_{0,0}(r,t) {Y_0}^{0,0} + \rho_{l,m}(r,t) {Y_0}^{l,m}~,~~~~l \ge 2 ~.
\end{equation}
The precise form of the $l=0$ mode $\rho_{0,0}$ is not important for the radiation outcome, 
since only $l \ge 2$ modes of the density do contribute to the right-hand-side of the
perturbation equation. In other words, the radiation is (at least) quadrupolar, so neither
the monopole nor the dipole part produce gravitational radiation. Notice however that,
although dynamically the $(0,0)$ mode is not important, it is the one that determines the total
mass of the matter distribution.

In this section, we start by presenting an study of a narrow Gaussian pulse of matter falling onto the
black hole and generating a gravitational wave, and show that we obtain the expected behavior, validating
in this way our equations and measuring the accuracy of our code. Next, 
we will consider three cases to study the gravitational response of the black hole to the
infalling matter. In the first case we study a single pulse
based on a gaussian distribution and will vary its width, describing the wave-form generated by the
corresponding pulses and measuring its dependence on the width of the initial pulse. In the second case, we will describe three consecutive gaussian pulses and describe
the gravitational radiation generated, in order to approach a model of matter falling onto the black hole in a
periodic manner. Finally, we present a third case with two consecutive Gaussian pulses, and vary the separation between them, in order to study the waveforms of the gravitational wave as a function of the infall frequency.

\subsection{Testing the code}

We start by considering initially a single Gaussian pulse of matter:
\begin{equation}\label{1prho_gaussian}
\rho_{l,m}(r,t=0) = {A_0} \,e^{-(r-r_0)^2/\sigma^2},
\end{equation}
where $A_0, r_0$ are the initial amplitude and the initial position of the center of the gaussian and $\sigma$ is its width. We will consider that the pulse is in rest at infinity, that is, $E=1$. We do not consider the gravitational perturbation generated by the fact that the pulse arrive to the finite position $r_0$, so at the initial time there is no gravitational radiation, that is  $R(t=0,r)=\psi(t=0,r)=\Pi(t=0,r)=0$.
Due to this inconsistency, a spurious radiation is developed very quickly from the pulse and infalls to the black hole. The scattered part goes away to infinity, leaving the pulse of matter moving with a consistent set of data. The initial pulse of dust is located at $r_0=70 M$, which is far enough in order to not interact strongly with the spacetime background until the initial spurious wave content left the domain.

\begin{figure}
\centering
\includegraphics[scale=0.4,angle=0]{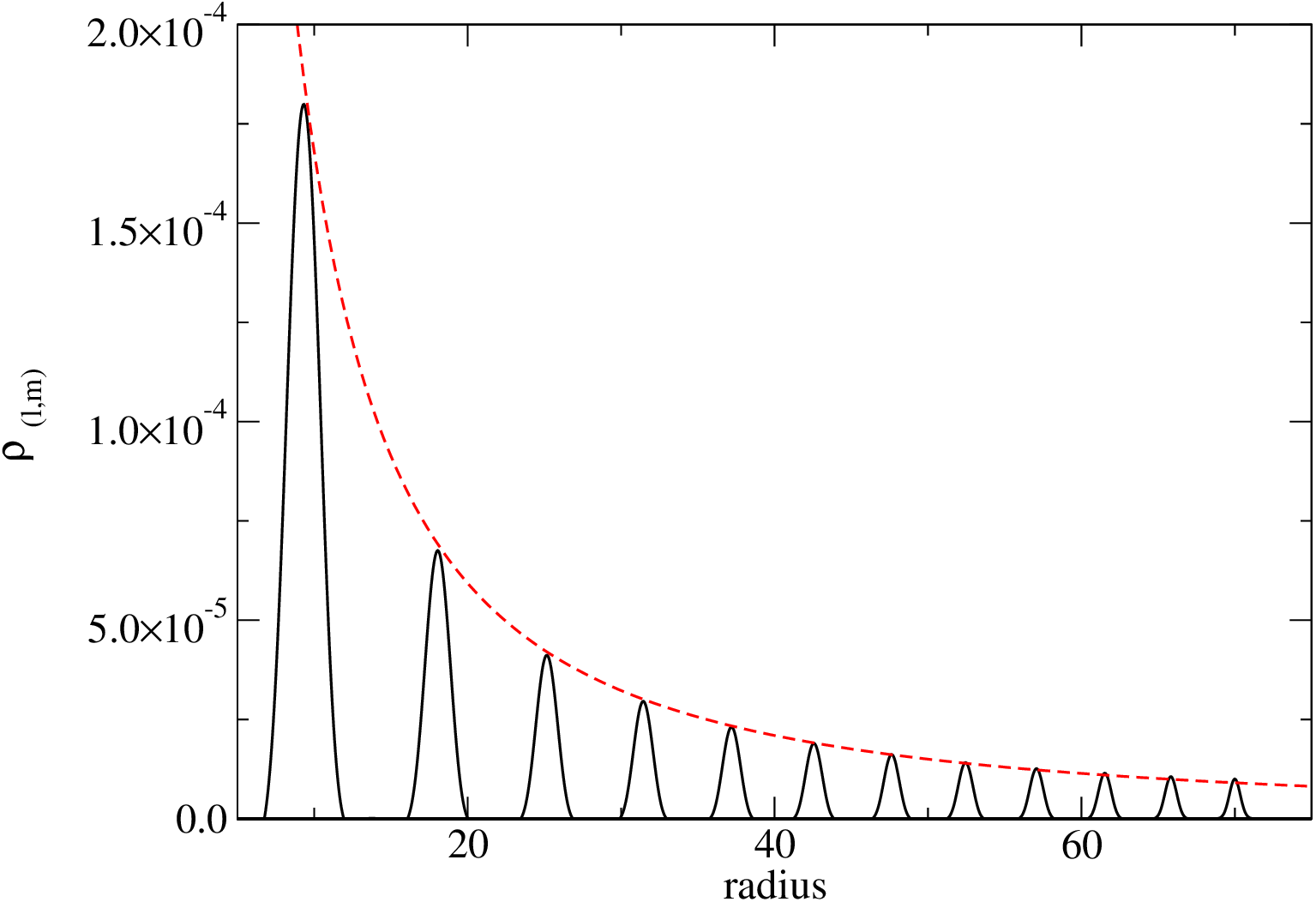}
\caption{The evolved density profile for a pulse of dust, plotted every $t=25M$, and the exact
envelope of the density. The gaussian has an initial amplitude of ${A_0}=10^{-5}$, an initial
width of $\sigma=0.5$ and is located at $r_0=70\,M$.}
\label{fig:pulse_evolution}
\end{figure}

\begin{figure}
\centering
\includegraphics[scale=0.4,angle=0]{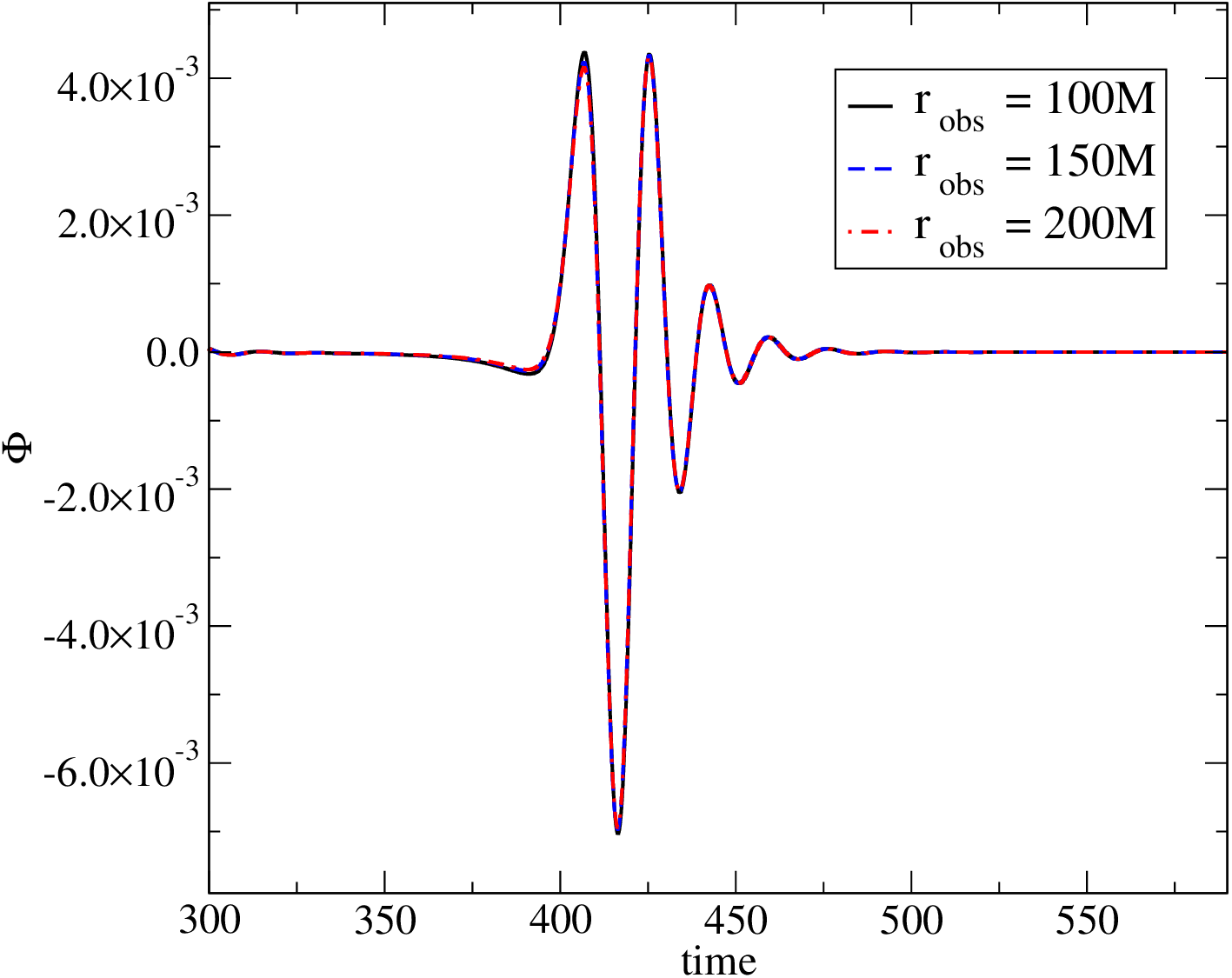}
\caption{Gravitational signal produced by the pulse from fig.~\ref{fig:pulse_evolution},
measured by three different observers located at $r=100\,M, 150\,M$ and $200\,M$. The signals
have been shifted in time in order to overlap them, as expected from the $1/r$ decay
of the $\Psi_4$.}
\label{fig:psi_observers}
\end{figure}

\begin{figure}
\centering
\includegraphics[scale=0.4,angle=0]{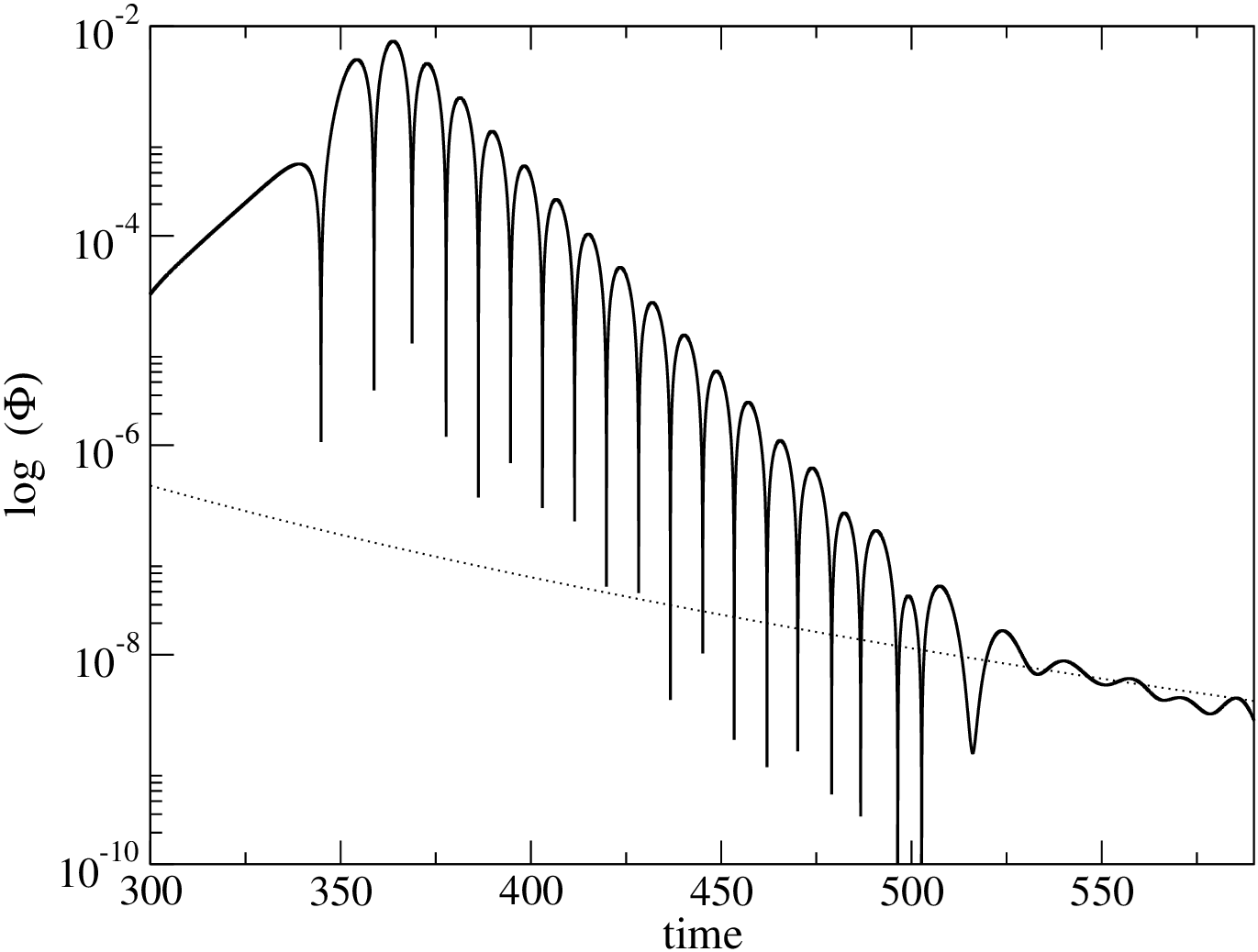}
\caption{The $\Phi$ from fig.~\ref{fig:psi_observers} in logarithmic scale (continuous line)
and a function proportional to $t^{-7}$ (dashed line). Notice how the late time behavior of the gravitational response (ie, the tail) decays like $t^{-7}$.}
\label{fig:sigma05}
\end{figure}

\begin{figure}
\centering
\includegraphics[scale=0.32,angle=0]{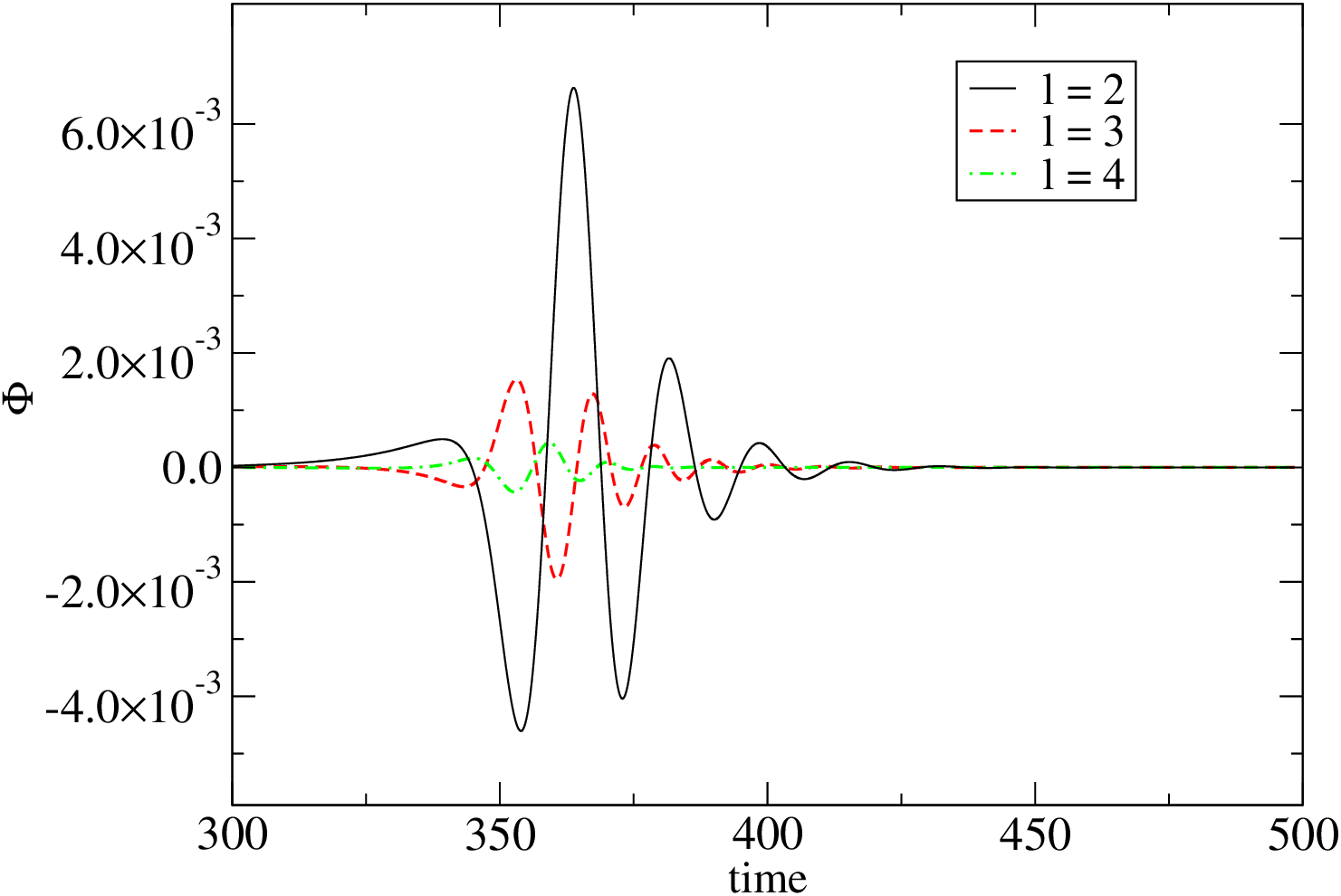}
\caption{Gravitational signal for different modes, namely $l=2, 3, 4$. All of them have the
same value for the initial amplitude, ${A_0}=10^5$, the same width, $\sigma=0.5$, and
start at the same position, $r_0=70\,M$.}
\label{fig:variasls}
\end{figure}

We assume first a density mode $\rho_{2,0}$, with an amplitude $A_0=10^{-5}$, and a width of $\sigma=0.5$.
In fig. \ref{fig:pulse_evolution} is compared the profile of such mode at different
times ({\it i. e.}, every $t=25 M$) with the exact solution for the envelope 
of the density
\begin{equation}\label{exact_rho}
    \rho_{l,m}(r) = A_0 \,\left(\frac{r_0}{r}\right)^\frac{3}{2}~~,
\end{equation}
which is computed by integrating the evolution equation (\ref{eq:evolrho})
for the mode $\rho_{l,m}$. It can be seen that the fluid is evolving properly since
the envelope fits during the evolution. 
In Fig.~\ref{fig:psi_observers} it is displayed the radiation $\Phi$ produced by this
shell of fluid, measured by three different observers placed at $r=100,150$ and $200 M$.
The signals has been shifted in order to overlap and stress the
expected $1/r$ behaviour of the $\Psi_4$ (ie, a constant behaviour of $\Phi$). From the same
signal we can adjust the ring-down form $e^{-\omega_I\,t}\,cos\left(\omega_r\,t\right)$
in order to obtain the quasinormal frequencies $\{\omega_R,\omega_I\}$. For this pulse, with
an observer located at $r_{\rm obs}=100\,M$, we obtain the values $M\,\omega=0.3734 - 0.0891\,i$, which are very close to the analytical values $M\,\omega=0.37367 - 0.08896\,i$ \cite{Leaver85, Kokkotas99}.
In fig.~\ref{fig:sigma05} it is displayed the same gravitational response $\Phi$ in logarithmic scale
in order to better appreciate the expected late time behavior, known as tail, which goes as $t^{-7}$ \cite{Pur05,Zeng08}.
In fig. \ref{fig:variasls} is shown the radiation produced by the same initial parameters for the profile $\rho_{l,m}$ for $l=2,3,4,5$. The corresponding QNM frequencies also are in good agreement with the theoretical ones \cite{Leaver85, Kokkotas99}, which are summarized in table I. 
\begin{table}\label{Tab:frecuencies}
\begin{center}

\caption{Quasinormal frequencies, notice that for the Schwarzchild case these values are independent of $m$}
\begin{tabular}[c]{l|l|l}
\hline \hline
$l$  & Computed $M\omega$& Analytical $ M\omega$ \\
\hline 
$2 $ & $0.3734 - 0.0891\,i$ & $0.37367 - 0.08896\,i$   \\
$3 $ & $0.5990-0.0932\,i$ &  $0.59944 - 0.09270\,i$ \\
$4 $ & $0.8055-0.0957\,i$ & $0.80918 - 0.09416\,i$ \\
\hline\hline
\end{tabular}

\end{center}
\end{table}

\subsection{One pulse with various widths}

\begin{figure}
\centering
\includegraphics[scale=0.65,angle=0]{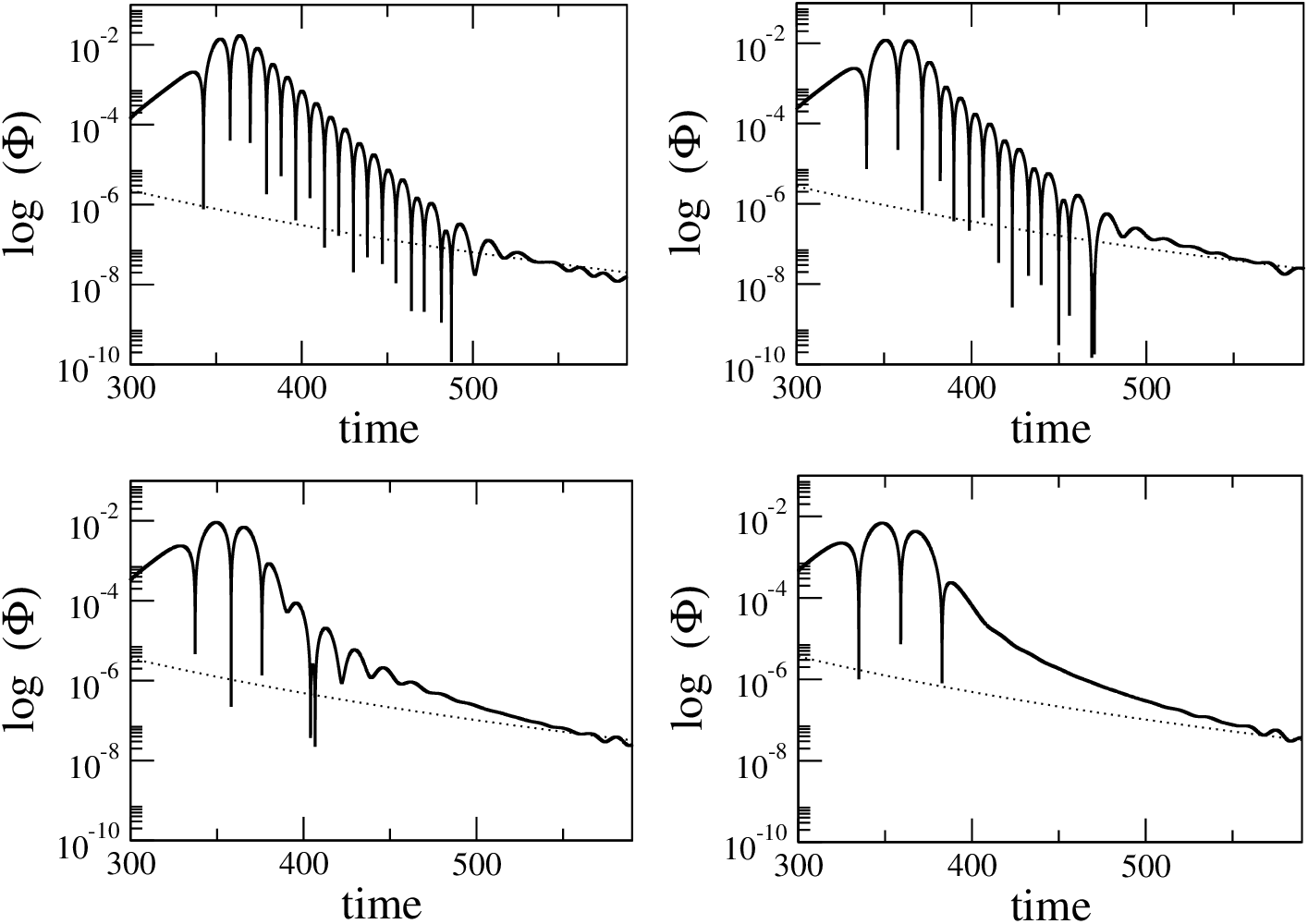}
\caption{The gravitational signal save for the different value of $\sigma$. We are plotting, from
left to right, for $\sigma=M, 1.5\,M, 2\,M, 2.5\,M$. Notice how the QNM frequencies become less noticeable, and the late time behavior differs from the theoretical one, as the matter pulse becomes wider. }
\label{fig:sigmas}
\end{figure}

\begin{figure}
\centering
\includegraphics[scale=0.4,angle=0]{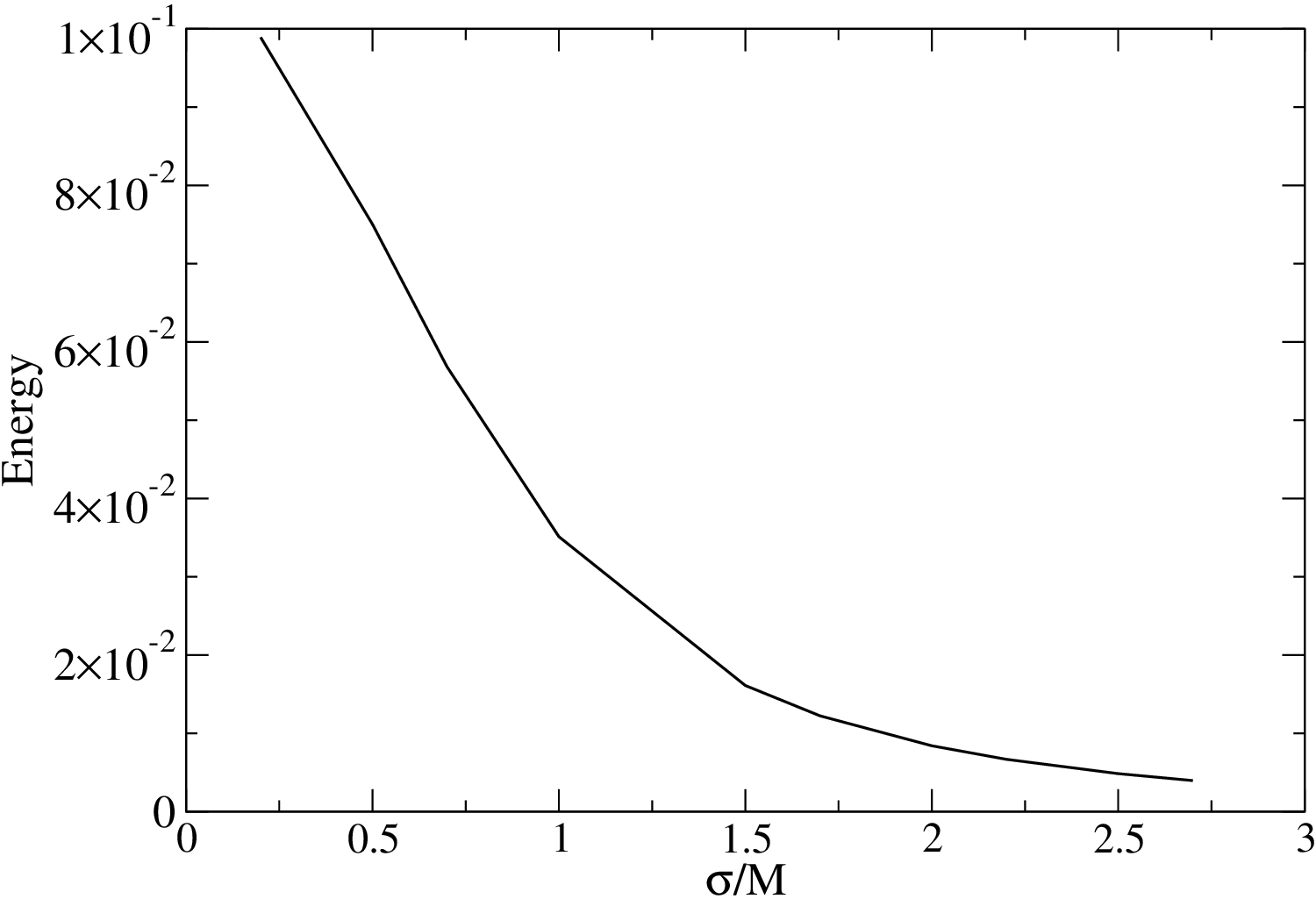}
\caption{Total energy as a function of the shell width, measured by an observer located at $r=70M$.}
\label{fig:energy}
\end{figure}

Once we have tested the accuracy of our code by comparing with already known results, 
we proceed to analyze the gravitational response for
different widths of the pulse. We studied the cases for $\sigma=0.5\,M, M, 1.5\,M, 2\,M$, and $2.5\,M$, and present the corresponding logarithmic plot of the gravitational response in fig.~\ref{fig:sigmas}. 
%
%
The outgoing energy flux per unit of time is given by:
\begin{equation}
\frac{dE}{dt}=\lim_{r \to \infty}\frac{r^{2}}{16\,\pi}\oint\Big | \int_{-\infty}^{t} \Psi_{4}\,dt'\Big |^{2}d\Omega 
  = \lim_{r \to \infty}\frac{1}{16\,\pi}\sum_{l,m}\Big | \int_{-\infty}^{t} R_{l,m}\,dt'\Big |^{2},
\label{energy_psi}
\end{equation}
where we used the multipole expansion (\ref{def:Phi_lm}) for $\Phi$ and the orthogonality of the spin weighted spherical harmonics. The total emitted energy can be computed by integrating in time the outgoing flux (\ref{energy_psi}), and the result for the different widths is displayed in
fig.~\ref{fig:energy} .\\
%
Our simulations show that the response of the black hole is the well know ring down only when the source of the perturbation acts in a very short period of time. As the source of the perturbation occurs during a longer interval of time, the signal loses the normal modes of the black hole and consequently the ring down behavior is gone. Also, with respect to the expected tail behavior, we see that the agreement with the theoretical exponential decay, gradually happens at later times as the matter signal is wider. These results have been also noticed by \cite{Nag07,Font98}, where they claim that ``the QNM excitation is induced by the curvature profiles that have spatial wave lengths comparable to the width of the black hole potential'', which is a conclusion similar
to our own.

Indeed we see that in general, the usual gravitational signal has an initial burst which is then followed by a decaying oscillating wave with the ring down frequencies. Our simulations show that as the pulse becomes wider the signal only keeps the first initial bursts. This behavior might have important consequences in the actual detection of gravitational waves as long as the ring down is not present in all the gravitational emissions of a perturbed black hole, and should be confirmed by simulations performed with data resembling realistic astrophysical scenarios.

\subsection{3 pulses}

Periodic and quasiperiodic variations are observed in different classes of astrophysical objects containing accretion discs. Quasiperiodic oscillations of thick accretion discs orbiting around a black hole have been addressed as sources of gravitational radiation \cite{Zanotti03}. In such cases, the tori oscillates induced by perturbations from the equilibrium configuration; as the inner ring of the tori approaches the black hole, part of its mass is moved through the cusp, resulting in a quasiperiodic accretion of matter onto the black hole. The dynamics of the tori is consequently printed in the gravitational wave signal.

In order to emulate this scenario we set up a density profile composed by three gaussian pulses, mimicing a periodic perturbation like the one expected in the accreting tori \cite{Nag07}.
Although we are not capturing the complete picture of the dynamics of the tori (like the angular momentum accreted), this approach allow us to study the main features of the gravitational signal produced by these periodic perturbations.

\begin{figure}
\centering
\includegraphics[scale=0.65,angle=0]{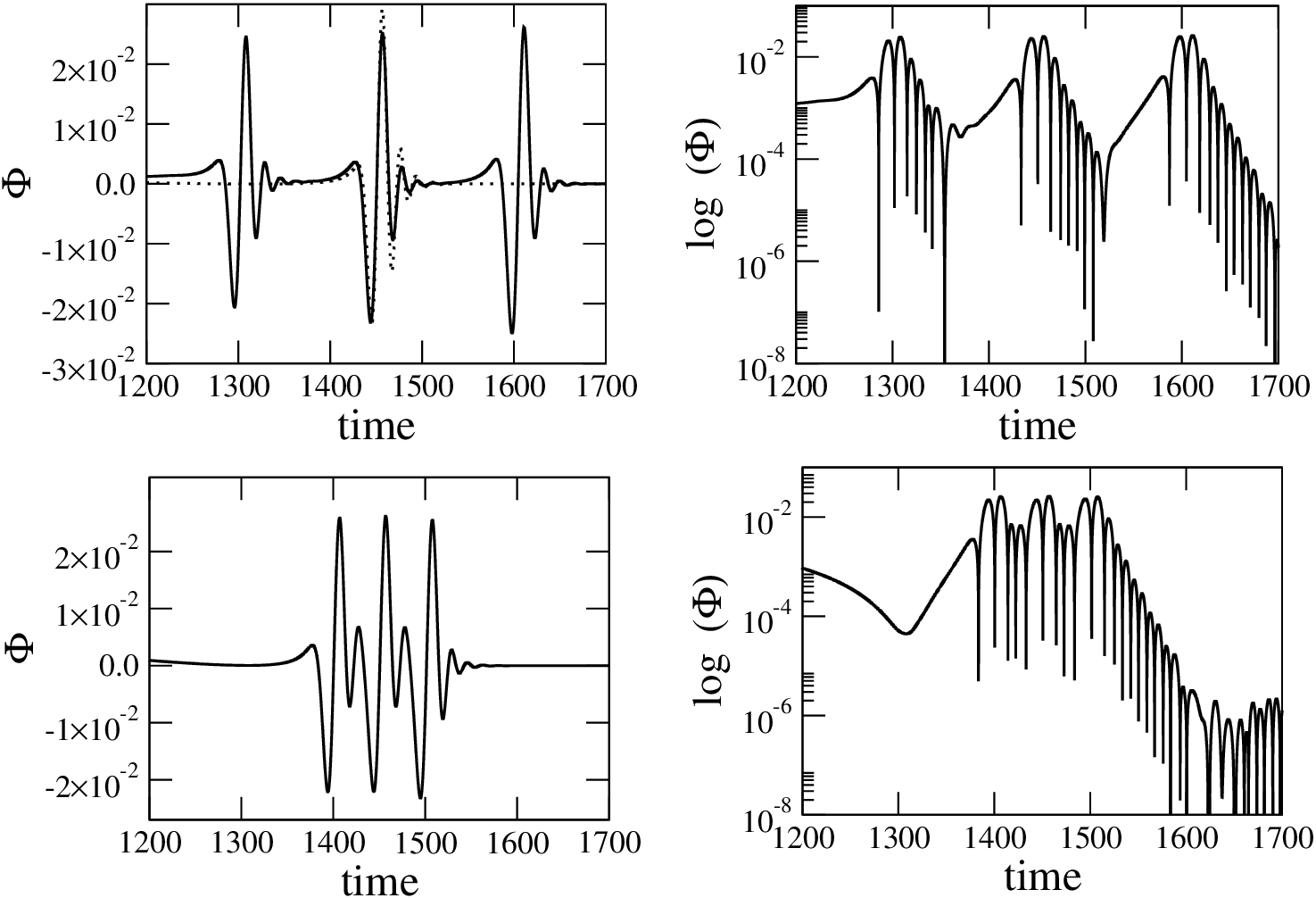}
\caption{Gravitational response due to the infall of three consecutive pulses. In the top panel we are showing the  $\Phi$ for three pulses with a separation $d=15M$, superposed with the response of one single pulse located at $r=200M$ in the left panel. In the bottom panels the separation between pulses has been reduced to $d=5M$.}
\label{fig:3pgrav}
\end{figure}

We analyze the gravitational response due to three consecutive pulses,
\begin{equation}\label{3prho_gaussian}
\rho_{2,0}(r,t=0) = A_0\,\left(e^{-(r-r_0)^2/\sigma^2} + e^{-(r-r_0 -d)^2/\sigma^2} 
                  + e^{-(r-r_0-2 d)^2/\sigma^2}\right).
\end{equation}
with the closest one located at $r_0=185 M$ and the other separated a distance $d=15M$ behind each other.
The initial amplitude is $A_0=1\times 10^{-5}$ and the width is $\sigma=0.5M$ for each of them.
We present in figure \ref{fig:3pgrav} (top panels) the gravitational signal measured by an observer located at $r=70M$, where the periodic signal of the initial density profile is printed upon it. Notice however how the ring down behavior of the gravitational response is preserved if the pulses are far enough from each other. To make clear this point we have superposed the signal that is produced by a single pulse located at $r_{0}=200M$
(ie, the central one) and notice that the general behavior coincides. This result differs from the one obtained by \cite{Nag07} in which they obtain that the gravitational response due to a periodic infall of matter generates a corresponding gravitational signal made of burst without the ring down behavior. This may happen when the 
infalling pulses are so close to each other that there is no time for the appearance of the normal modes in the gravitational responses, as it is shown in fig \ref{fig:3pgrav} (bottom panels) ,where the separation between pulses is just $d = 5M$. This behaviour will be analyzed in more detail in the next
subsection.

\subsection{2 pulses}

\begin{figure}
\centering
\includegraphics[scale=0.65,angle=0]{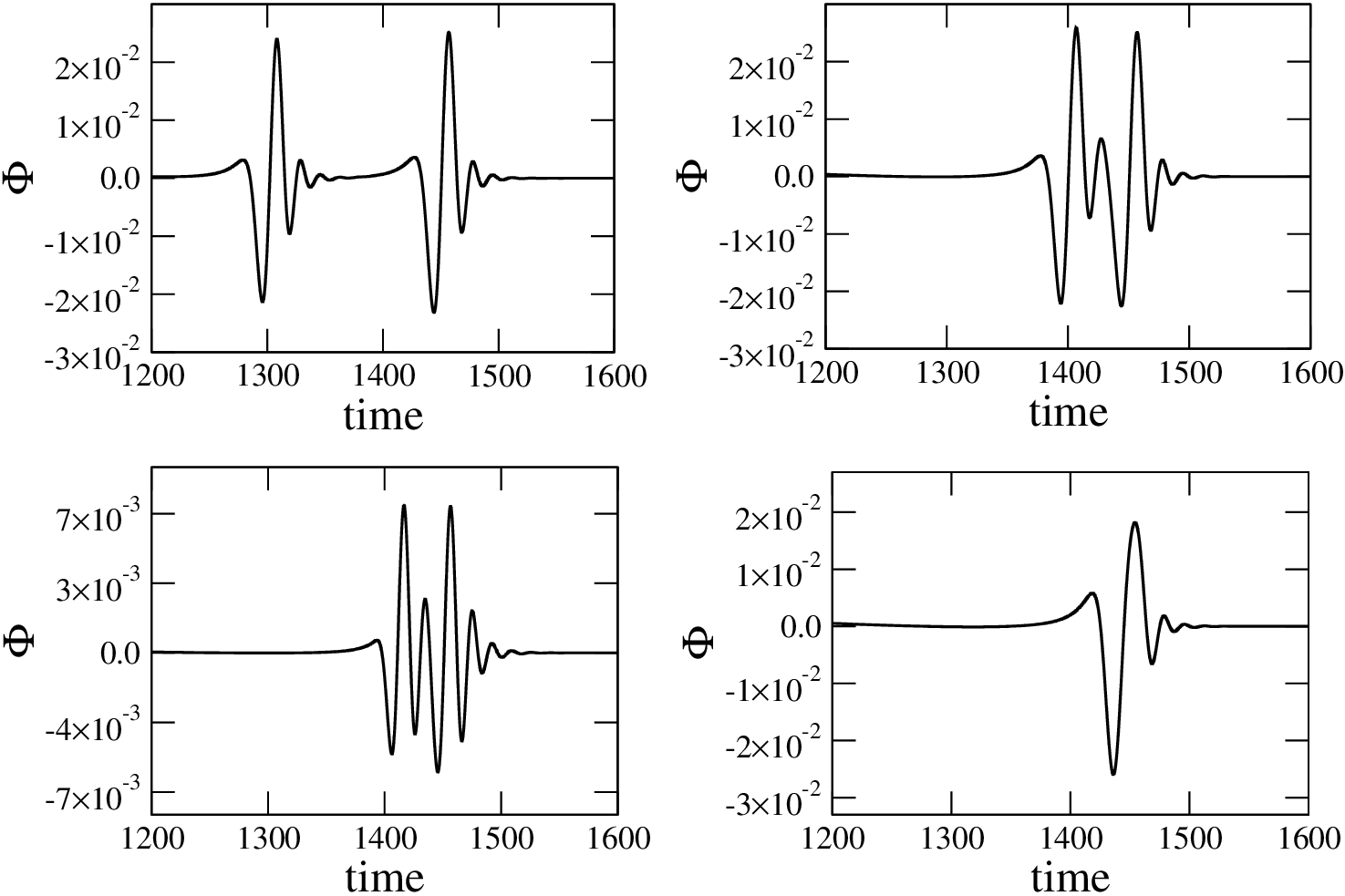}
\caption{It is shown the gravitational response of two consecutive pulses varying the initial separation $d$ between them. These separations are, from left to right an up down, $d=15M$, $d=5M$, $d=4M$ and $d=1M$. 
}
\label{fig:2psepar}
\end{figure}

In this section we consider the case of two consecutive pulses, varying the separation between them,
in order to study in detail the transition between a ring-down response of the black hole and a forced 
oscillation due to the infall of quasiperiodic pulses. The initial data is a density profile of the form:
\begin{equation}\label{2prho_gaussian}
\rho_{2,0}(r,t=0) = A_0\,\left(e^{-(r-r_0)^2/\sigma^2} + e^{-(r-r_0 - d)^2/\sigma^2}\right).
\end{equation}
where $d$ varies between $[15-1]$. The center of the first pulse is $r_{0}=200$ and we use
$\sigma=0.5M$ as usual. In figure \ref{fig:2psepar} it is shown how the signal mirrors the behavior of the shell that is falling into the black hole; as the separation between the pulses decreases, the time elapsed between the signals measurement also decreases, up to the point where the observer is unable to distingish between two consecutive signals. This happens in the second and third panel of the plot, where each signal has lost their individuality superposing each other, to form a signal that hides the ring down behavior of the individual.
Finally in the fourth panel the pulses are initially so close $1M$ that the gravitational response of
the hole is the same as produced by a single perturbation with the double of the width. 
Based on these results we can conclude that the initial separation of the matter pulses actually affects the gravitational response of the black hole. In particular, the signal corresponding to the pulse arriving first when the second one is close ($5M$) has no  time to generate a ring down signal.
The last pulse however does present the ring down behavior as expected.
In particular when the initial separation is $5\,M$ the signal resembles a succession of bursts similar to the ones obtained by \cite{Nag07}. Finally, we notice that when the pulses are even closer to each other the gravitational response becomes the one due to a single wider pulse. Regarding the separation of consecutive pulses, there is a similar behavior as in the case of the width of the pulse. The gravitational signal is affected when the initial 
separation between the pulses is comparable to the radius of the black hole.

\section{Discussion} \label{sec:dis}

We have rederived the evolution equation for gravitational perturbations including sources, showing how to deal in a consistent manner with different choices of signature. As a first application, we have studied the spherically symmetric static spacetimes, in particular the Schwarzschild black hole. After
some comparison with previous similar works, we have constructed the tetrad and choose the penetrating coordinates to compute the perturbation equation with matter source terms.

By means of the expansion of the rest mass density in spherical harmonics, we were able to describe a generic distribution of dust radially falling onto the black hole as a one dimensional problem.
We have implemented the perturbation and the matter equations numerically to study the gravitational response in several interesting cases of the infalling dust. We have first discussed the importance of the width in the infalling shells in the generated gravitational signal, showing how the ringdown behavior is lost in favor of a single burst response, when the pulse becomes wider.
Afterwards we have studied the case of infall of consecutive pulses. Our claim is that this is a way to study the gravitational perturbation of a black hole, when the matter is released in a periodic fashion, as in an oscillating tori. Indeed, we are thinking on a tori which initially fills a region just down to the Roche's lobe and due to some perturbation, starts to oscillate and in each period, some of its matter gets further than the Roche's lobe, consequently falling onto the black hole and producing a periodic source of gravitational perturbation. We have modeled the periodic emission of matter by the tori as consecutive pulses of radially infalling dust. Moreover, once the matter passed beyond the Roche's lobe, its motion could be consider is practically radial, so our approach allow us to focus in this stage of the motion. There were claims that in these cases of periodic infall of matter, the gravitational response was modified losing the ring down behavior in favor of a response dominated by the frequency of the infalling matter as in a forced motion, we started with three consecutive pulses and obtained a gravitational signal with three burst each one with the spected ring down behavior, with no further effect of the periodicity of the infalling matter.
Finally, we proceded to study the gravitational response of two consecutive infalling pulses varying the initial separation and obtained that indeed there is an interval of the separation of the pulses in which the gravitational response is certainly affected, clearly when there is no time for the ring down to happend due to the infall of the next pulse, however if the initial separation gets even shorter the gravitational response is that of a single wider pulse.
In this way we have showed that the gravitational signal is not only determined by the parameters of the space time it moves but it has also imprints of the manner in which it has been generated   

The present work gives a solid base and sets us in a track that should be follow on other cases, like the one with matter or a small compact body falling onto a rotating black hole. The presence of certain types of pressure can give rise to oscillating motions of spherical shells, as discussed in \cite{Nun97, Nun98}, which could generate a continue source of perturbation to the black hole, which in turn might generate a continued emission of gravitational waves, carrying information of the background as well as of the matter which produced them.

\section*{Acknowledgments}
We are greatfull to Luciano Rezzolla, Sasha Hussa, Anil Zengino\u{g}lu, Cecila Chirenti, Shin Yoshida, and Jos\'e A. Font, for many useful discussion during the elaboration of this work. D N\'u\~{n}ez is greatfull to Luciano Rezzolla for warm hospitality during his stay at AEI, and acknowledges DAAD, DGAPA-UNAM and CONACyT grants for partial support. JCD acknowledges the CONACyT schoolarship.

\bibliographystyle{unsrt}
\bibliography{refs}
 \end{document}